\documentclass{aa}
\usepackage[varg]{txfonts}   % A&A strongly recommends this for proper layout
\usepackage{graphicx}
\usepackage{amsmath}
\usepackage{booktabs}
\usepackage{threeparttable}
\usepackage[caption=false]{subfig}
\usepackage{epstopdf}        % keep if you really need .eps files
\usepackage{bm}
\usepackage{natbib}
\usepackage{placeins}
\usepackage[pdfauthor={Yuping Tang et al.}, 
            colorlinks=true,
            linkcolor=blue,
            citecolor=blue,
            urlcolor=blue,
            pdfborder={0 0 0}]{hyperref}
\bibpunct{(}{)}{;}{a}{}{,}

\begin{document}

\title{Effects of low resolution on the column density PDF of molecular clouds}

\author{Yuping Tang\inst{1}
\and Q.~Daniel Wang\inst{2}
\and Grant W.~Wilson\inst{2}
\and Xing Lu\inst{3}
\and Sihan Jiao\inst{4}
\and Jinhua He\inst{5,6,7}}

\institute{
  \inst{1} Shanghai Key Lab for Astrophysics, Shanghai Normal University, Shanghai 200234, People's Republic of China \\
  \inst{2} Department of Astronomy, University of Massachusetts Amherst, Amherst, MA 01002, USA \\
  \inst{3} Shanghai Astronomical Observatory, Chinese Academy of Sciences, 80 Nandan Road, Shanghai 200030, People's Republic of China \\
  \inst{4} Max Planck Institute for Astronomy, Königstuhl 17, D-69117 Heidelberg, Germany \\
  \inst{5} Yunnan Observatories, Chinese Academy of Sciences, Kunming 650216, China \\
  \inst{6} Chinese Academy of Sciences South America Center for Astronomy, National Astronomical Observatories, CAS, Beijing 100101, P.R. China \\
  \inst{7} Departamento de Astronom\'{i}a, Universidad de Chile, Las Condes, 7591245 Santiago, Chile
}

\abstract{
  {Observational resolution significantly impacts the interpretation of column density probability distribution functions (N-PDFs) in molecular clouds, which are essential for understanding turbulent structures and star formation processes.}
  {This study quantifies how low spatial resolution truncates the high-density power-law tails of N-PDFs by simulating distant observations (2--10 kpc) of 17 local massive molecular clouds/regions using Herschel-based column density maps.}
  {We propose a parameter-free model, assuming proportional embedding of dense regions within lower-density gas, to predict the truncation column density where the survival function equals the beam-to-threshold area ratio.}
  {Comparisons with simulations show good agreement, with deviations up to 0.3 dex attributed to cloud multiplicity in large complexes and flatter power-law tails in coherent structures. Characteristic cloud scales, derived from $\Delta$-variance spectra, indicate that global smearing dominates when beam sizes exceed these scales. We further develop a reverse method to recover the intrinsic high-density tail from low-resolution data.}
  {Our findings link N-PDF shapes to morphologies, suggesting that feedback-compressed extended structures resist smearing, while multiplicity accelerates truncation. These insights caution against biases in N-PDF decompositions and provide a framework for correcting resolution effects in distant cloud studies, enhancing constraints on star formation theories.}
}


\maketitle
\nolinenumbers

\section{Introduction} \label{sec:intro}

Star formation is essentially a hierarchical structure formation process spanning multiple orders of magnitude in spatial scales \citep{maclow04,krumholz14}. Theoretical models need to accurately describe the multi-level collapse process from molecular clouds to protostellar cores, which poses extremely high requirements on the precision of numerical simulations and the coupling of physical processes. As a key statistical measure of turbulent gas structures, the volume density-probability density distribution function ($\rho$-PDF) of molecular clouds not only reflects multi-scale turbulent structures but is also used as a core criterion for verifying numerical simulations \citep{vazquez94,kowal07}. Current mainstream star formation theories \citep{krumholz05,mckee07,hennebelle11,burkhart18} mostly use $\rho$-PDF as the basic initial condition, calculate the star formation rate by integrating its high-density tail (combined with the collapse critical density threshold), and introduce timescale factors and efficiency factors for correction. Due to technical obstacles in directly measuring $\rho$-PDF, observational studies mainly obtain the column density probability distribution function (N-PDF) through means such as infrared-millimeter band dust emission, near-infrared extinction, and molecular line tracers \citep{kainulainen09,lombardi15,chen18,schneider15,schneider22,spilker21}. Under simple geometric assumptions (i.e., a spherical shape with a power-law density profile, Girichidis et al. 2014), the N-PDF can be used to derive the corresponding $\rho$-PDF form through an integral transformation, while recent advancements have extended this process with more sophisticated reconstruction techniques that relax simplistic assumptions and enhance accuracy for diverse molecular cloud morphologies (e.g., Orkisz \& Kainulainen 2025). Constraining the N-PDF of molecular clouds through observations provides key constraints for understanding the cross-scale evolution of star formation in turbulent media.

It is generally believed that the $\rho$-PDF/N-PDF has two typical distribution intervals: at the low-density end, turbulence and random shock processes lead to a log-normal distribution. The physical mechanism originates from random fluctuations in density within turbulent gas, which are dominated by multiplicative processes conforming to the statistical laws of the central limit theorem \citep{vazquez94}. Numerical simulation studies \citep{kritsuk11,federrath13} indicate that the standard deviation of the molecular cloud $\rho$-PDF/N-PDF in the log-normal interval is positively correlated with the turbulent Mach number and is significantly influenced by the turbulent driving mode \citep{padoan97,federrath08} and magnetic field strength \citep{molina12}. At the high-density tail, the gravity-dominated self-similar collapse process forms a power-law distribution. The analytical study of \citet{girichidis14} proves that, after undergoing incompressible collapse, free-falling molecular clouds evolve a power-law tail in their PDF within a time shorter than the free-fall timescale, with the power-law index of the $\rho$-PDF being -1.54. Assuming a spherical geometric structure and a power-law density profile, this power-law index can be transformed into the power-law index of the N-PDF as -2.11.

Observational studies indicate that the actual form of the molecular cloud N-PDF is more complex than theoretical expectations. First, the correlation between the standard deviation of the log-normal region of the N-PDF predicted by theory and the turbulent Mach number has so far lacked strong observational support \citep{price11,goodman09,kainulainen17}. Second, large-sample studies \citep{schneider15,schneider22,spilker21,ma22} have also found that the N-PDF may exhibit a mixed form of log-normal-power-law distribution, and the power-law distribution may also present multi-segment structures, with the latter's possible triggering mechanisms being magnetic fields or stellar feedback.

Observational studies of N-PDFs often face three challenges: 1. The low-density end of the N-PDF is contaminated by foreground and background emissions along the line of sight; 2. observations of local molecular clouds may suffer from limited coverage due to their large angular extent; 3. limited spatial resolution can lead to systematic underestimation of the standard deviation and truncation of power-law tails. In particular, recent studies have extended research on N-PDFs to distant clouds \citep{battersby25,jiao25}, highlighting the need for a systematic understanding of resolution effects on N-PDFs. While the effects of low resolution on the N-PDF, which primarily cause truncations at both high and low density ends, have been briefly discussed in some works \citep{ossenkopf16,alves17}, so far, there is no quantitative, systematic study on the location of the truncation column density in the N-PDF. 

In this paper, we aim to understand how low-resolution effects influence our interpretation of the N-PDFs of molecular clouds. The primary objective is to identify the principal factors influencing the truncation behavior of N-PDFs at high densities. This analysis is based on a sample of local, massive clouds, for which low-resolution images are simulated as if observed from large distances, allowing the observed N-PDFs to be treated as intrinsic. The subsequent step involves developing a model to replicate the truncation behavior in these simulated low-resolution images, while evaluating its successes and limitations. We will demonstrate that such a model requires only the intrinsic N-PDF of the cloud and its area as inputs to effectively predict the truncation behavior of N-PDFs. Furthermore, we will show that these results enable the construction of a framework to recover the ``missing" high-density tail resulting from low resolution.

This paper is organized as follows. The sample collection and data reduction are described in Section~\ref{sec:sample}. In Section~\ref{sec:model}, we propose a simple model based on the assumption of proportional nesting to predict the truncation behavior of clouds' N-PDFs, demonstrate the model's robustness, and discuss the outliers and their implications. In Section~\ref{sec:recover}, we show how this model can be reversed to reconstruct the missing high-density tail of the N-PDFs from low-resolution observations. Section~\ref{sec:discuss} provides discussions. In Section~\ref{sec:summary}, we draw our conclusions.

\section{Sample and data preprocessing}\label{sec:sample}

\subsection{Sample selection and data reduction}

To understand how N-PDFs evolve with distance and resolution, we constructed a sample of 17 regions within local (d $<$ 1 kpc) massive molecular clouds that have available Herschel data or column density maps, as listed in Table~\ref{tab:cloud}. This distance limit for the sample clouds ensures that we have their high-resolution N-PDFs, which will be treated as their intrinsic N-PDFs. Most regions have column density map products inferred from dust Spectral Energy Distribution (SED) fits available from the data archive of the Herschel Gould Belt Survey (HGBS; \citep{andre10}); we also obtained the column density maps of the California cloud from the HP2 (Herschel-Planck) survey \citep{lada17}. Additionally, we included Mon OB1 West, Mon OB1 East, Mon R2, and Vela C in our sample, which are not in the HGBS survey. For these four regions, the Herschel 160--500 $\mu$m maps were downloaded from the Herschel Science Archive. For the SPIRE 250--500 $\mu$m data, the level 2.5 product maps were downloaded, which had already been corrected for zero-point offsets. For the PACS 160 $\mu$m data, the level 2.0 product maps were downloaded, and the zero-point offset was corrected by utilizing the 140 $\mu$m map from the AKARI Far-infrared All-Sky Survey, through first fitting an SED to FIS/AKARI 140 $\mu$m + SPIRE/Herschel 160-500 $\mu$m maps, degrading every map to the lowest 36.3'' resolution at SPIRE 500$\mu$m, and then interpolating the flux density at 160 $\mu$m. Column densities were derived by fitting 160-500 $\mu$m dust SEDs, assuming a $\beta$ of 2 and $\kappa_0$ = 0.1 cm$^2$/g at $\nu_0$ = 1000 GHz, similar to the approach adopted by the HGBS \citep{roy14}. One exception is the column density map of the California cloud, which was derived by assuming a free $\beta$ and a fixed $\nu_0$ at 353 GHz. However, as will be shown later, our model is not sensitive to the absolute calibration uncertainties of the column densities. The SED fitting was performed using a Bayesian analysis approach with slice/Gibbs sampling (Tang et al. 2021, Section 4.1 and Appendix A). The only difference from \citet{tang21} is that we conduct pixel-by-pixel fitting throughout this work, after degrading all maps to the lowest 36.3'' resolution.

For each cloud, we set a column density threshold corresponding to a continuous closed contour, which defines the ``effective" area to be studied. The thresholds are generally chosen according to two principles: 1. If the cloud complex (e.g., Orion B, which can be divided into L1630 N and L1630 S; Perseus, which can be divided into Perseus East and Perseus West) can be divided into sub-regions, each enclosed by a continuous closed contour, contour thresholds for the sub-regions are used; 2. If the cloud has a high background due to low Galactic latitudes, higher thresholds are used to separate the background contribution. A background level is measured for each cloud, typically by averaging two low-density regions, one above the cloud and one below the cloud relative to the Galactic Plane. The contour thresholds and background levels are also listed for each cloud in Table~\ref{tab:cloud}.

\begin{table*}[]
\centering
\caption{Clouds/Regions Information}
\label{tab:cloud}
\resizebox{\textwidth}{!}{
\begin{threeparttable}
\begin{tabular}{lcccccc}
\toprule
Cloud/Region & Distance [pc]\tnote{a} & $\log_{10}(N_{\rm thresh}/{\rm cm}^{2})$\tnote{b} & $\log_{10}(N_{\rm bk}/{\rm cm}^{2})$\tnote{c} & Area [pc$^2$]\tnote{d} & $M_{\odot}$\tnote{e} & L$_{\rm char}$ [pc]\tnote{f} \\
\hline
MonOB1 W & 699 & 21.60 & 21.23 & 25.9 & 3139 & 0.25 \\
MonOB1 E & 699 & 21.60 & 21.23 & 26.7 & 4504 & 0.47 \\
MonR2 & 836 & 21.50 & 21.08 & 18.7 & 2572 & 1.03 \\
VelaC N & 990 & 21.75 & 21.38 & 41.9 & 7504 & 0.43 \\
VelaC SC & 990 & 21.75 & 21.38 & 258.3 & 50671 & 0.34 \\
California & 527 & 21.25 & 20.98 & 447.8 & 16993 & 0.21 \\
OrionA & 388 & 21.20 & 20.48 & 264.4 & 24458 & 0.57 \\
L1630 N & 412 & 21.35 & 20.60 & 26.2 & 3171 & 0.49 \\
L1630 S & 412 & 21.35 & 20.60 & 47.7 & 4509 & 0.22 \\
Perseus E & 319 & 21.35 & 20.78 & 19.0 & 1334 & 0.10 \\
Perseus W & 289 & 21.25 & 20.78 & 17.7 & 1446 & 0.16 \\
Serpens M & 436 & 21.50 & 21.23 & 35.9 & 2768 & 0.24 \\
Serpens S & 440 & 21.80 & 21.49 & 53.8 & 8594 & 0.29 \\
Serpens NE & 478 & 21.55 & 21.40 & 39.1 & 2036 & 0.20 \\
Cepheus1241 & 342 & 21.05 & 20.70 & 33.7 & 831 & 0.27 \\
Taurus L1495 & 138 & 21.30 & 20.90 & 8.9 & 595 & 0.08 \\
IC5146 & 740 & 21.15 & 20.78 & 43.1 & 2498 & 0.51 \\
\hline
\end{tabular}
\begin{tablenotes}[flushleft]
\small
\item[a] The distance values are taken from \citet{zhang23}.
\item[b] Logarithmic threshold column density (for defining the contour lower limit of clouds/regions).
\item[c] Logarithmic background column density.
\item[d] The area enclosed by a continuous, closed contour exceeding $  N_{\rm thresh}  $ in the original high-resolution map.
\item[e] Dust inferred total gas mass above $N_{\rm thresh}$.
\item[f] Characteristic size of clouds/regions (defined as the turning point of the 
$\Delta$-variance spectrum).

\end{tablenotes}
\end{threeparttable}
}
\end{table*}

\subsection{Simulating low-resolution maps at large distances}\label{sec:mimic}

The low-resolution column density maps for each region are simulated at five distances: $  d =  $ 2, 4, 6, 8, and 10 kpc. Rather than scaling the image inversely with increasing distance, this simulation is achieved by first degrading the resolution through convolution of the original column density map with a Gaussian beam of standard deviation $\sigma_{\rm cov} = \sqrt{(\sigma_{\rm beam})^2 - \sigma_0^2}  $, where $\sigma_{\rm beam} = \sigma_0 \cdot r_{\rm d}$ represents the standard deviation of the apparent beam in the simulated maps.$  \sigma_0 = 36.3'' / (2 \sqrt{2 \ln 2})  $ and $  r_{\rm d} = d / d_0  $ (with $  d_0  $ denoting the original distance to the region). The convolved image is then rebinned to a pixel size of $  3'' \times r_{\rm d}  $ to preserve a consistent sampling rate relative to the beam size.

To ensure a fair comparison between the column density probability density functions (N-PDFs) derived from the low-resolution maps and those from the original high-resolution maps, we employed the following strategy:\\
\\
1) For each low-resolution map, define a continuous, closed contour based on the threshold values $  N_{\rm thresh}  $ listed in Table~\ref{tab:cloud}, and use it to generate a corresponding mask. \\
\\
2) Project this contour back onto the high-resolution map by mapping its boundary coordinates to the finer grid and including all high-resolution pixels that lie within or on the boundary, thereby creating an equivalent high-resolution mask.\\
\\
3) In the high-resolution mask, set the values of all pixels below the specified threshold to zero. Note that the region defined by the mask from step 2) typically includes pixels exceeding the threshold values, owing to the generally more porous nature of high-resolution maps. Across the sample, the ratios of the high-resolution mask areas to those of the low-resolution masks range from 0.5 to 1.\\

\begin{figure*}[!htbp]
\centering
\includegraphics[scale=0.8]{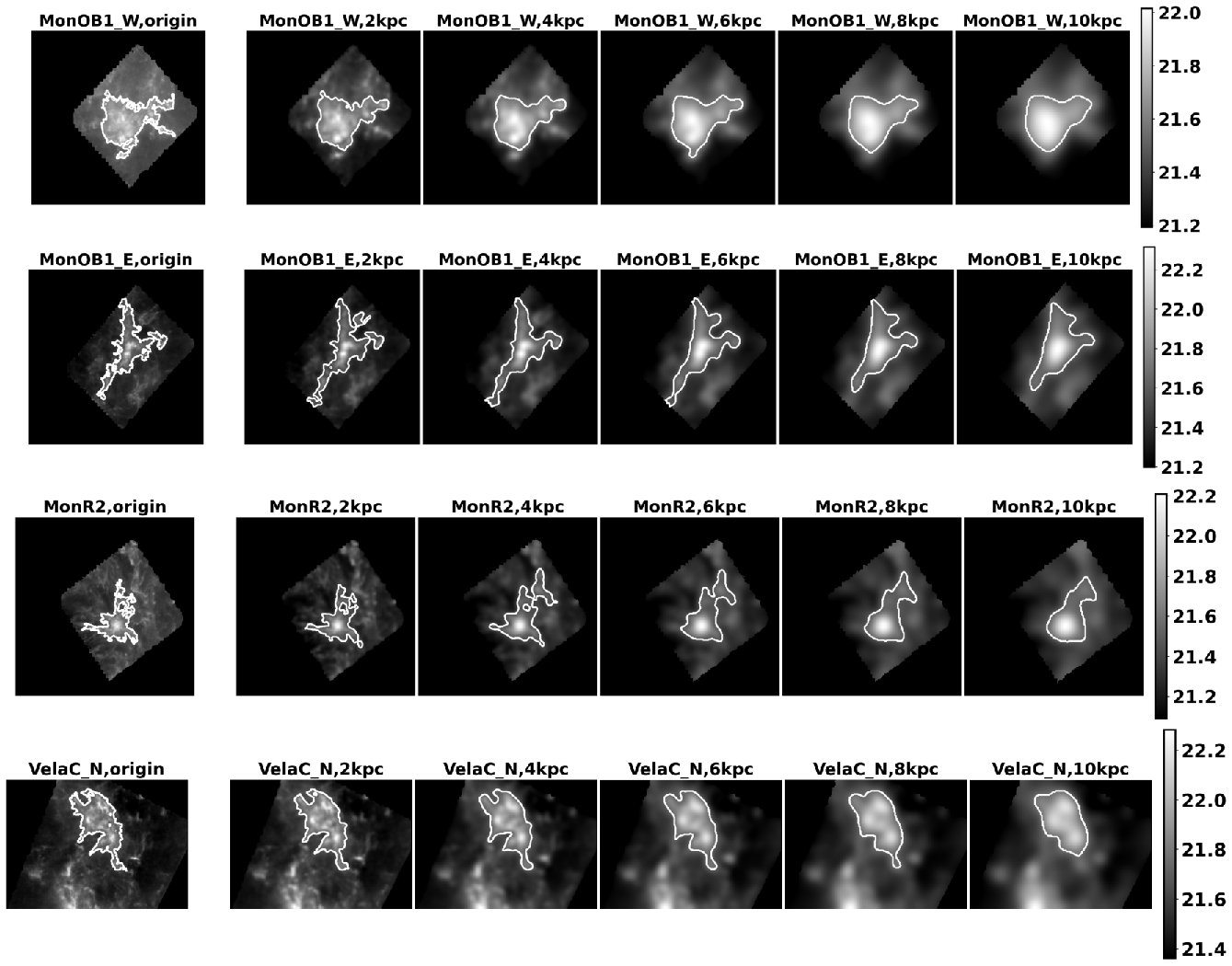}
\caption{Examples of simulated low-resolution images of the clouds and regions at different distances. For each cloud/region, the original high-resolution map is shown in the leftmost panel, followed by the simulated maps at 2, 4, 6, 8, and 10 kpc. The continuous closed white contour in each panel corresponds to the threshold column density \(N_{\rm thresh}\) listed in Table~\ref{tab:cloud}. See the Appendix for plots of the entire sample.}
\label{fig:simu}
\end{figure*}

\begin{figure*}[!htbp]
\centering
\includegraphics[scale=0.45]{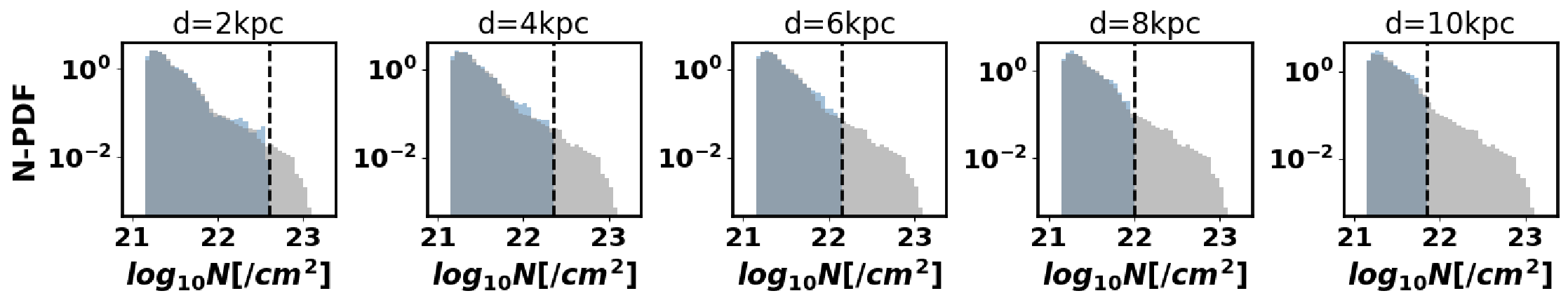}
\caption{Comparisons between low-resolution and high-resolution N-PDFs for the Serpens Main region. The vertical dashed line marks $  \log_{10}N_{\rm trun, sim}$ calculated by Equation~\eqref{eq:ntrun}.}
\label{fig:example}
\end{figure*}

The simulated low-resolution images are presented in Figure~\ref{fig:simu}. Figure~\ref{fig:example} also shows an example of the corresponding N-PDFs at low and high resolutions for each distance. The truncation in the N-PDFs attributable to low resolution exhibits a clear negative correlation with distance. The target areas range from 9 to 448 pc$  ^2  $, and the dust-inferred total masses enclosed within these areas range from 595 to 50,671 $  M_{\odot}  $. Consequently, this sample spans a sufficiently wide range in terms of morphology and mass, enabling us to quantify the effects of low resolution on the N-PDFs.

\section{Proportional nesting model for truncation column density}\label{sec:model}

\subsection{Proportional nesting}\label{sec:nest}

N-PDF, by definition, represents the probability density of observing a given column density $  N  $, and is proportional to the fraction of pixels (or area) with that density. Consequently,

\begin{equation}\label{eq:nest}
P(>N) \propto A(>N)
\end{equation}

where $  P(>N)  $ is the cumulative probability for column densities exceeding $  N  $, and $  A(>N)  $ is the corresponding area fraction above this threshold.

Consider a simple, nested density structure within a molecular cloud, where higher-density regions (e.g., cores or clumps) are embedded or enclosed within lower-density envelopes in a contiguous, connected hierarchy. This configuration implies a monolithic geometry without significant disconnected components, analogous to a single submerged mountain, where the peak (representing the dense core) emerges as a solitary entity before expanding into broader slopes (lower-density structures) as the ``water level'' (density threshold) decreases. For such a structure, we further have:

\begin{equation}\label{eq:model}
P(>N_{\rm trun}) = A_{\rm beam}/A_{\rm thresh}
\end{equation}

where:

\begin{equation}
A_{\rm beam} = 2\pi\sigma_{\rm beam}^2
\end{equation}

and $  A_{\rm thresh}  $ is the angular area enclosed by the threshold contour defining the region of interest. 

The simple nesting described above is rarely found in molecular clouds. Consider an extreme case in which two identical clumps with the same N-PDF are analyzed as a single structure. In such a scenario, $N_{\rm trun}$ would be overestimated according to Equation~\eqref{eq:model}, as $A_{\rm thresh}$ is doubled while $P(>N)$ remains unchanged from that of a single clump. In turbulent clouds, however, density fluctuations follow self-similar statistics, where multiplicity manifests as multiple compact substructures embedded within a larger diffuse envelope. Unlike multiple isolated identical clouds; Multiplicity is turbulence-driven, characterized by clumps that are connected via filaments, shared envelopes, or velocity-coherent structures and spatially clustered. This means $A_{\rm thresh}$ and $P(>N)$ are not independently scalable—the substructures contribute to a self-similar hierarchy where high-N regions are proportionally nested within lower-N gas, even if visually clumpy. As long as these substructures are smaller than the beam size at low resolutions and their collective area distribution aligns with the high-resolution $P(>N)$, beam smearing dilutes them proportionally, mimicking the predicted truncation, and proportionality~\eqref{eq:nest} is still largely retained. We call this assumption proportional nesting.

To test the extent to which Equation \eqref{eq:model} holds in reality, we assume that the intrinsic N-PDF of each region can be represented by its observed N-PDF, $P_{\rm high}$. We measure $N_{\rm trun,sim}$ from simulated maps as the column density at which the low and high resolution probability density ratio $P_{\rm low}(\log_{10}N_{\rm trun,sim}) / P_{\rm high}(\log_{10}N_{\rm trun,sim})$ drops to $1/e$. Throughout this work, we compute the probability density function $  P(\log_{10} N)  $ from the images, rather than $  P(N)  $. These are related by:

\begin{equation}\label{eq:plgN}
P(\log_{10} N) = N \ln(10) \, P(N)
\end{equation}

which follows from the conservation of probability: $  P(\log_{10} N) \, d \log_{10} N = P(N) \, dN  $. In practice, $P_{\rm low}(\log_{10}N)$ and $P_{\rm high}(\log_{10}N)$ are represented by normalized histograms of the region enclosed by the threshold contour in the high-resolution map and low-resolution maps, respectively, with a bin width $\Delta_{\log_{10}N}$ of 0.05. $N_{\rm trun,sim}$ is given by:

\begin{equation}\label{eq:ntrun}
\log_{10}N_{\rm trun, sim} = \frac{\rm bin_{ cen}[i_{max}] + \rm bin_{ cen}[i_{max}+1]}{2}
\end{equation}

where $\rm bin_{cen}[i_{ max}]$ is the maximum bin center of the histograms satisfying $P_{\rm low}[{\rm i_{max}}] / P_{\rm high}[{\rm i_{max}}] > 1/e$. The $  \log_{10} N_{\rm trun,sim}  $, as defined, is illustrated in Figure~\ref{fig:example}.

\begin{figure}[!htbp]
\centering
\includegraphics[scale=0.3]{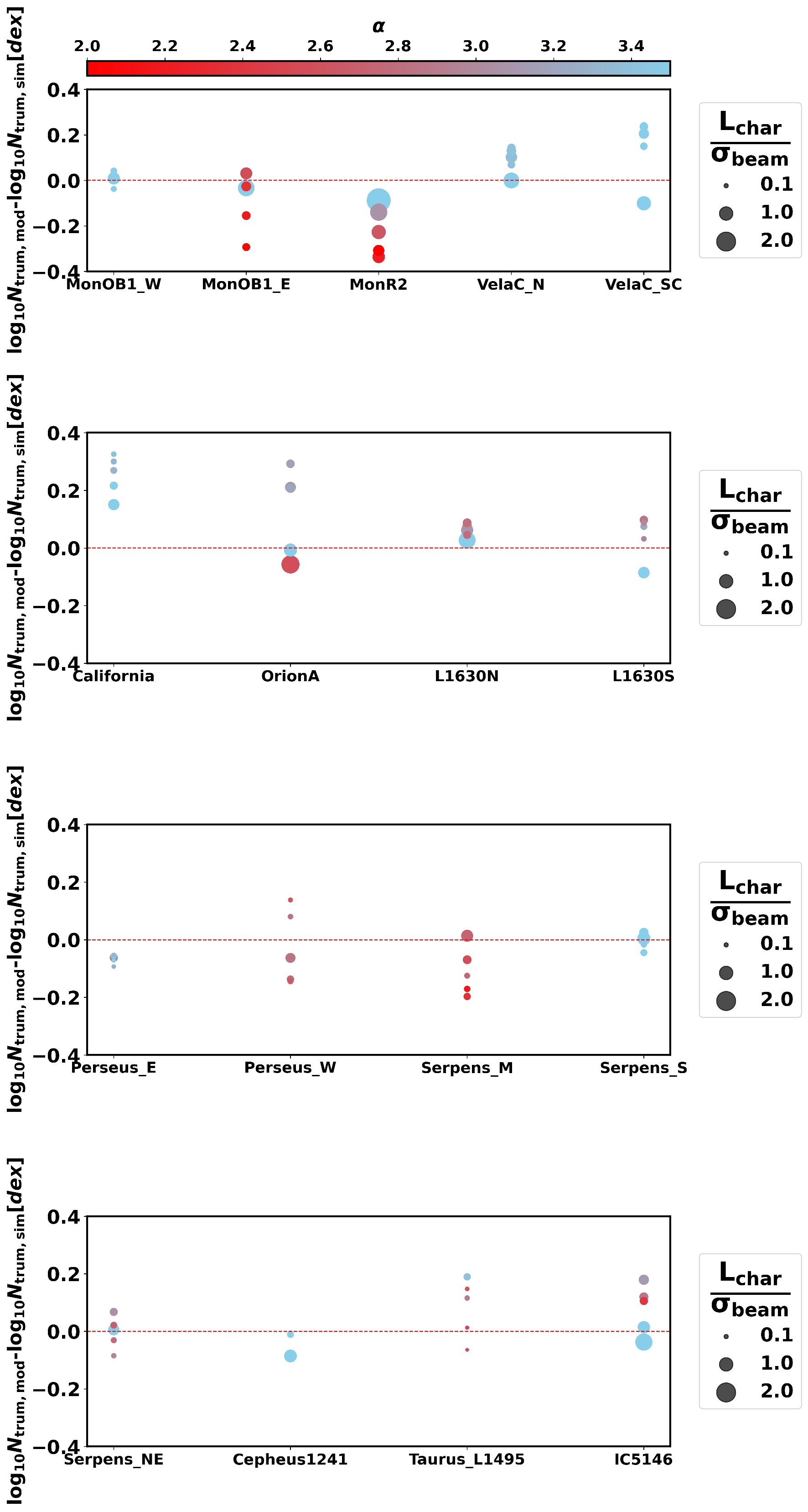}
\caption{Differences $\log_{10}(N_{\rm trun,mod}) - \log_{10}(N_{\rm trun,simu})$ between model-predicted and simulated truncation column densities for all clouds/regions at distances of 2--10\,kpc. Colors indicate the SFEq power-law index derived from the high-resolution N-PDF above $N_{\rm trun,simu}$; symbol sizes show the ratio of characteristic cloud size to beam size.}
\label{fig:ratios}
\end{figure}

The ratios of the model-predicted $N_{\rm trun, mod}$ by equation~\eqref{eq:model} to the measured $N_{\rm trun,sim}$ for all regions are summarized in Figure~\ref{fig:ratios}. Overall, $\log_{10}(N_{\rm trun, mod}) - \log_{10}(N_{\rm trun,sim})$ is close to 0, with deviations of up to 0.3 dex. This indicates a remarkable agreement between our simple model based on proportional nesting and the simulations, especially considering that the model has no free parameters.

\subsection{Regions with overestimated $N_{\rm trun,mod}$}

Three large complexes in our sample—Orion A, Vela C South \& West, and California—exhibit most overestimated $N_{\rm trun,mod}$, with $\Delta \log_{10}(N_{\rm trun,mod}/N_{\rm trun,sim})>0.2$ dex at large distances. This is naturally expected based on the discussion in Section~\ref{sec:nest}. These complexes feature continuous structures, making it difficult to separate them into isolated regions by simply increasing the threshold column density. As a result, the chosen contour covers hundreds of pc² in area, which likely suffers significantly from multiplicity and consequently leads to overestimated $N_{\rm trun,mod}$. However, even for these regions, the overestimated ratio of approximately 2 (or 0.3 dex) is still significantly smaller than that expected for parallel structures within the region, thereby supporting our assumption of proportional nesting.

\subsection{Regions with underestimated $N_{\rm trun,mod}$}

Contrary to the model overestimation discussed in Section 3.2, an opposing class of model underestimation relative to the simulation is linked to the shape of the high-density tail itself. It is observed that the regions exhibiting the most underestimated $  N_{\rm trun, mod}  $, including Mon R2, Mon OB1 East, and Serpens Main, also display a flat power-law tail above $  N_{\rm trun}  $ in their high-resolution N-PDFs. To quantify the potential relationship between underestimated $  N_{\rm trun, mod}  $ and a flat power-law tail, we define a survival-function-equivalent (hereafter SFEq) power-law index $  \alpha_{\rm SFEq}  $, derived under the assumption that the integral of a power law from $  N_{\rm trun}  $ to infinity equals that of the high-resolution N-PDF over the same range:

\begin{equation}\label{eq:sfeq}
\int_{N_{\rm trun}}^{\infty} P_{\rm high}(N_{\rm trun})(\frac{N}{N_{\rm trun}})^{-\alpha} dN = \int_{N_{\rm trun}}^{\infty} P_{\rm high}(N) dN
\end{equation}

Combining Equation~\eqref{eq:sfeq} with Equation~\eqref{eq:plgN}, one can derive $  \alpha_{\rm SFEq}  $ as:

\begin{equation}\label{eq:alpha1}
\alpha_{\rm SFEq} = 1 + \frac{P_{\rm high}(\log_{10}N_{\rm trun})}{\ln(10)P_{\rm high}(>\log_{10}N_{\rm trun})}
\end{equation}

Smaller $\alpha_{\rm SFEq}$ values correspond to a heavier dense-gas tail.

In Figure~\ref{fig:ratios}, each data point is color-coded according to the SFEq power-law index of the cumulative N-PDF above $  N_{\rm trun}  $. This confirms that the aforementioned regions—Mon R2, Mon OB1 East, and Serpens Main—possess the smallest $  \alpha_{\rm SFEq}  $, indicating that regions with the flattest power-law tails are less susceptible to beam smearing than anticipated from the proportional nesting model. The implications of this correlation will be discussed in Section~\ref{sec:discuss}.

\begin{figure*}
\centering
\includegraphics[scale=0.7]{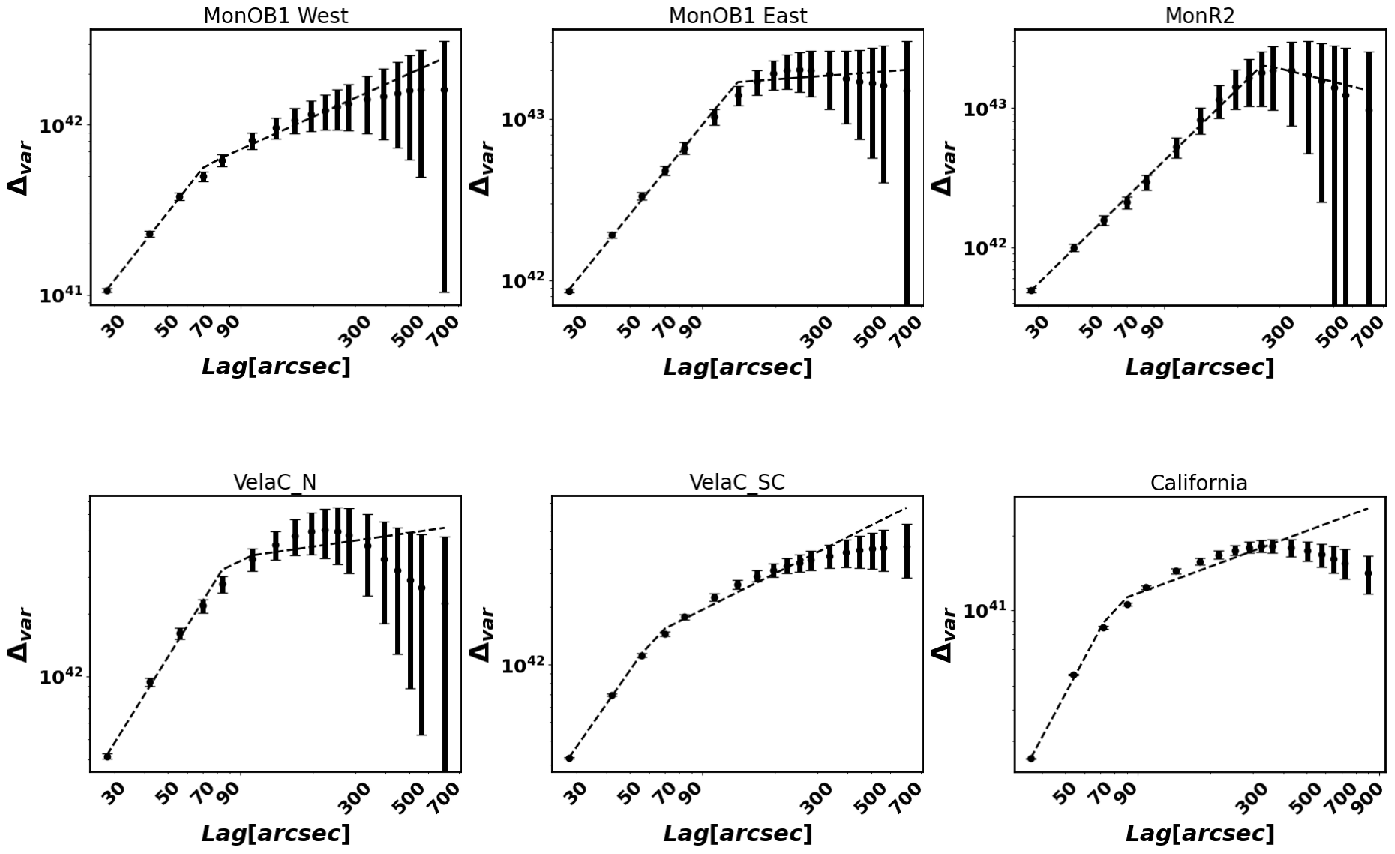}
\caption{Examples of $\Delta$-variance spectra of the sample clouds/regions, computed using a Mexican-hat filter \citep{ossenkopf08a,ossenkopf08b}. Dotted lines show the best-fit broken power laws. The smallest lag included is $3\times$ FWHM to avoid resolution effects. See the Appendix for plots of the entire sample.}
\label{fig:deltav}
\end{figure*}

\subsection{Characteristic sizes of cloud regions}

In real observations that incorporate a degree of multiplicity, one might anticipate a characteristic scale threshold at which beam smearing transitions from local (within substructures) to global (encompassing multiple substructures and intervening gaps), thereby amplifying deviations from the ideal proportional nesting assumption in clouds with enhanced multiplicity. A reasonable estimate for this scale threshold is the point at which power-law-like self-similar behavior breaks down in the clouds; such a scale could be provided by the $\Delta$-variance spectrum of individual clouds.

The $\Delta$-variance spectrum is a wavelet-based tool that quantifies the spatial scaling of intensity variations in column density maps (Stutzki et al. 1998, Ossenkopf et al. 2008a \& b). It is computed by convolving the map with a filter (e.g., a Mexican hat wavelet) of varying lag size and calculating the variance as a function of that scale. A $\Delta$-variance spectrum is similar to a power spectrum---which refers to the Fourier transform-based representation of the spatial variance distribution in a column density map--but robust to finite map sizes and edge effects. It typically shows a power-law rise $\Delta_{\rm var}(L) \propto L^\beta$, where $L$ is lag scale. The power index, is related to the power index of the power spectrum, denoted as $  \gamma  $ where $  P(k) \propto k^{-\gamma}  $, by the formula $  \beta = \gamma - 2  $. A turning point or ``break'', which marks the characteristic scale $L_{\rm char}$ of dominant structures, could be identified from the spectrum (e.g., transition from turbulent to self-gravitating regimes, or clump/filament correlation lengths ~0.1–1 pc in typical molecular clouds). \citet{dib20} demonstrate that the turning point in the delta-variance spectrum represents the characteristic size of a molecular cloud by constructing synthetic column density maps that overlay discrete 2D Gaussian structures—modeling over-dense features like clumps, filaments, and hubs—onto fractal Brownian motion (fBm) backgrounds with a power-law power spectrum, reflecting self-similar turbulence. Following this work, \citet{schneider22} employed $\Delta$-variance in analyzing N-PDFs across 29 Galactic clouds, finding that the visual extinction ($A_V$) at which the N-PDF slope changes between power-law tails increases with the characteristic size scale identified in the variance spectrum, linking turbulent mixing at low densities to self-gravity dominance at higher ones.

Following Ossenkopf et al. (2008a \& b), we use the ``Mexican hat'' filter to calculate the $\Delta$-variance of an image with a certain scale $l$. The filter $\oplus_l(\mathbf{r})$ at scale $l$ is composed of a positive core and a negative annulus, each normalized to unit integral:

\begin{equation}
\oplus_l(\mathbf{r}) = \oplus_{l,\text{core}}(\mathbf{r}) - \oplus_{l,\text{ann}}
(\mathbf{r})
\end{equation}

For the Mexican-hat filter:

\begin{equation}
\begin{split}
\oplus_{l,\text{core}}(\mathbf{r}) &= \frac{1}{4\pi (l/2)^2} \exp\left( -\frac{|\mathbf{r}|^2}{(l/2)^2} \right)
\end{split}
\end{equation}

\begin{equation}
\begin{split}
\oplus_{l,\text{ann}}(\mathbf{r}) &= \frac{1}{4\pi (l / 2)^2 (v^2 - 1)} \\
&\quad\left[ \exp\left( -\frac{|\mathbf{r}|^2}{(vl / 2)^2} \right) - \exp\left( -\frac{|\mathbf{r}|^2}{(l/2)^2} \right) \right]
\end{split}
\end{equation}

where $v=1.5$. We further employ the mask map corresponding to the threshold column density listed in Table~\ref{tab:cloud} as the weighting map. The zero-padded high-resolution map and the weighted map are then convolved with the Mexican hat filter, combined, and renormalized to generate the final $\Delta$-variance map (Equation 7-9 in \citet{ossenkopf08b}). The $\Delta$-variance spectra for all cloud regions are shown in Figure~\ref{fig:deltav}. We fit these spectra using a broken power-law function and derive the turning point as the characteristic scale of each cloud region, $L_{\rm char}$, as shown by Figure~\ref{fig:deltav}. The smallest scale included in the fit is 3 $ \sigma_{\rm beam}  $, to avoid attenuation arising from insufficient resolution at small lags $  L  $. 

As shown in Figure~\ref{fig:ratios}, the proportional nesting model accurately predicts $N_{\rm trun}$ when $L_{\rm char}/\sigma_{\rm beam} \lesssim 1$, even for large complexes. However, in the latter cases, the discrepancy $\Delta \log_{10}N$ increases to 0.2--0.3 dex when the beam size is $\gtrsim 2 \times$ the characteristic size. This can be interpreted as a scale threshold at which beam smearing shifts from local (within substructures) to global (encompassing multiple substructures and gaps), thereby amplifying non-uniform effects in multiplicity-enhanced clouds. At high resolutions, where beam smearing dilutes high-$N$ peaks locally while preserving the overall N-PDF tail, the perturbation remains minor; the proportional nesting model (based on ensemble area fractions) holds even under mild multiplicity, as the beam does not yet mix across disconnected components or large voids. In contrast, when the beam size exceeds this threshold relative to $L_{\rm char}$, characteristic structures become unresolved. This global smearing becomes non-uniform owing to enhanced multiplicity, with high-$N$ regions scattered amid gaps, resulting in more aggressive redistribution of densities.

\section{Recovering high density N-PDF from low-resolution observations}\label{sec:recover}

The proportional nesting model assumes that the truncation column density $  N_{\rm trun}  $ in the observed column density probability density function (N-PDF) of molecular clouds arises from beam smearing when the survival function $  P(>N)  $ equals the ratio of the beam area to the area of high-density structures. This model has demonstrated strong empirical agreement in predicting $  N_{\rm trun}  $ across simulated distances for nearby clouds. This success validates the model's underlying assumption of a hierarchical density structure, where dense regions are embedded within larger, lower-density envelopes, and smearing effectively redistributes high-$  N  $ material below observational thresholds.
Based on this success, it becomes possible to infer an SFEq power-law index $  \alpha  $ for the intrinsic high-$  N  $ tail—even from low-resolution data where the tail is truncated—by utilizing the model's quantification of the ``missing'' cumulative probability $  P(>N_{\rm trun})  $ beyond $  N_{\rm trun}  $. By anchoring a power-law extrapolation to the observed PDF value near $  N_{\rm trun}  $ and integrating it to match $  P(>N_{\rm trun})  $. Combining Equation~\eqref{eq:alpha1} with Equation~\eqref{eq:model}, we have:

\begin{equation}\label{eq:alpha2}
\alpha = 1 + \frac{P_{\rm high}(\log_{10}N_{\rm trun})A_{\rm thresh}*C_{\rm fill}}{\ln(10)A_{\rm beam}}
\end{equation}

Here, $  C_{\rm fill}  $ denotes a correction factor that accounts for the observation that the area exceeding a given threshold $  N_{\rm thresh}  $ is larger when measured at low resolution than at high resolution, owing to a smaller filling factor at higher resolutions. $  C_{\rm fill}  $ is defined as the mean value of $  A_{\rm high}/A_{\rm low}  $ across the 17 regions, and it ranges from 0.9 to 0.7 as the distance $  d  $ increases from 2 kpc to 10 kpc. Equation~\eqref{eq:alpha2} provides an approximation of the intrinsic tail's slope that accounts for resolution biases.

In Equation~\eqref{eq:alpha2}, $  P_{\rm high}(\log_{10}N_{\rm trun})  $ represents the intrinsic N-PDF in log scale evaluated at $  \log_{10}N_{\rm trun}$, which cannot be derived directly from low-resolution observations in theory. In practice, $  P_{\rm high}(\log_{10}N_{\rm trun})  $ can be approximately inferred by fitting a power law to the apparent tail of the N-PDF (spanning from the knee just below $  \log_{10}N_{\rm trun}$ to a selected low-N point). Subsequently, $  \log_{10}N_{\rm trun}  $ is defined as the density at which the observed N-PDF significantly deviates from this extrapolated power law $P_{\rm low}(\log_{10}N_{\rm trun}) = 1/e \times  P_{\rm extp}(\log_{10}N_{\rm trun})$. Then we have:

\begin{equation}\label{eq:alpha3}
P_{\rm high}(\!\log_{10\!}N_{\rm trun}) = P_{\rm extp}(\!\log_{10}\!N_{\rm trun}\!)[1\!-\!P(\!>\!\log_{10}\!N_{\rm trun})]
\end{equation}

The factor $1-P(>\log_{10}N_{\rm trun})$ comes from the approximation:

\begin{equation}
\begin{split}
\frac{P_{\rm extp}(\log_{10}N_{\rm trun})}{P_{\rm high}(\log_{10}N_{\rm trun})} \approx \frac{P_{\rm low}(\le\log_{10}N_{\rm trun})}{P_{\rm high}(\le\log_{10}N_{\rm trun})}  \\ 
\approx \frac{1}{1-P_{\rm high}(>\log_{10}N_{\rm trun})}
\end{split}
\end{equation}

\begin{figure*}[!htbp]
\centering
\includegraphics[scale=0.42]{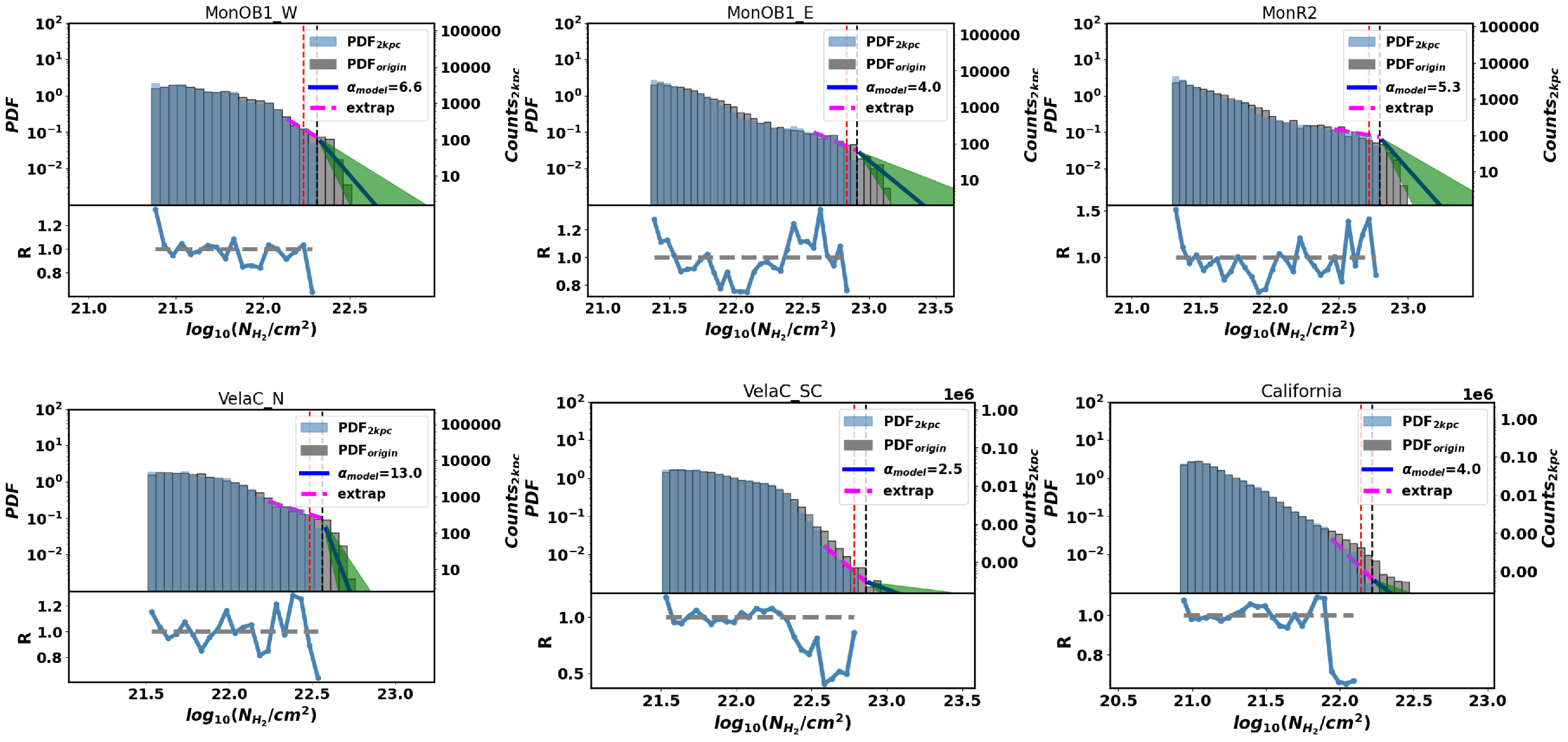}
\caption{Examples of recovering the power-law tail from simulated low-resolution observations at 2 kpc. Upper Panel: The blue line indicates the SFEq power-law estimated from Equation~\eqref{eq:alpha1}. The green shades indicate an uncertainty of $50\%$. The magenta line shows the best-fit power-law for extrapolation to derive $ P(\log_{10}N_{\rm trun}) $. Lower Panel: The ratio between the low-resolution PDF and the high-resolution PDF: $R=\frac{PDF_{\rm 2kpc}}{PDF_{\rm origin}}$. The black vertical line marks the observed $N_{\rm trun}$ derived from equation~\eqref{eq:ntrun}, the red vertical line marks the first point of maximum curvature. See the Appendix for the plots of the entire sample.}
\label{fig:2kpc}
\end{figure*}

\begin{figure*}[!htbp]
\centering
\includegraphics[scale=0.42]{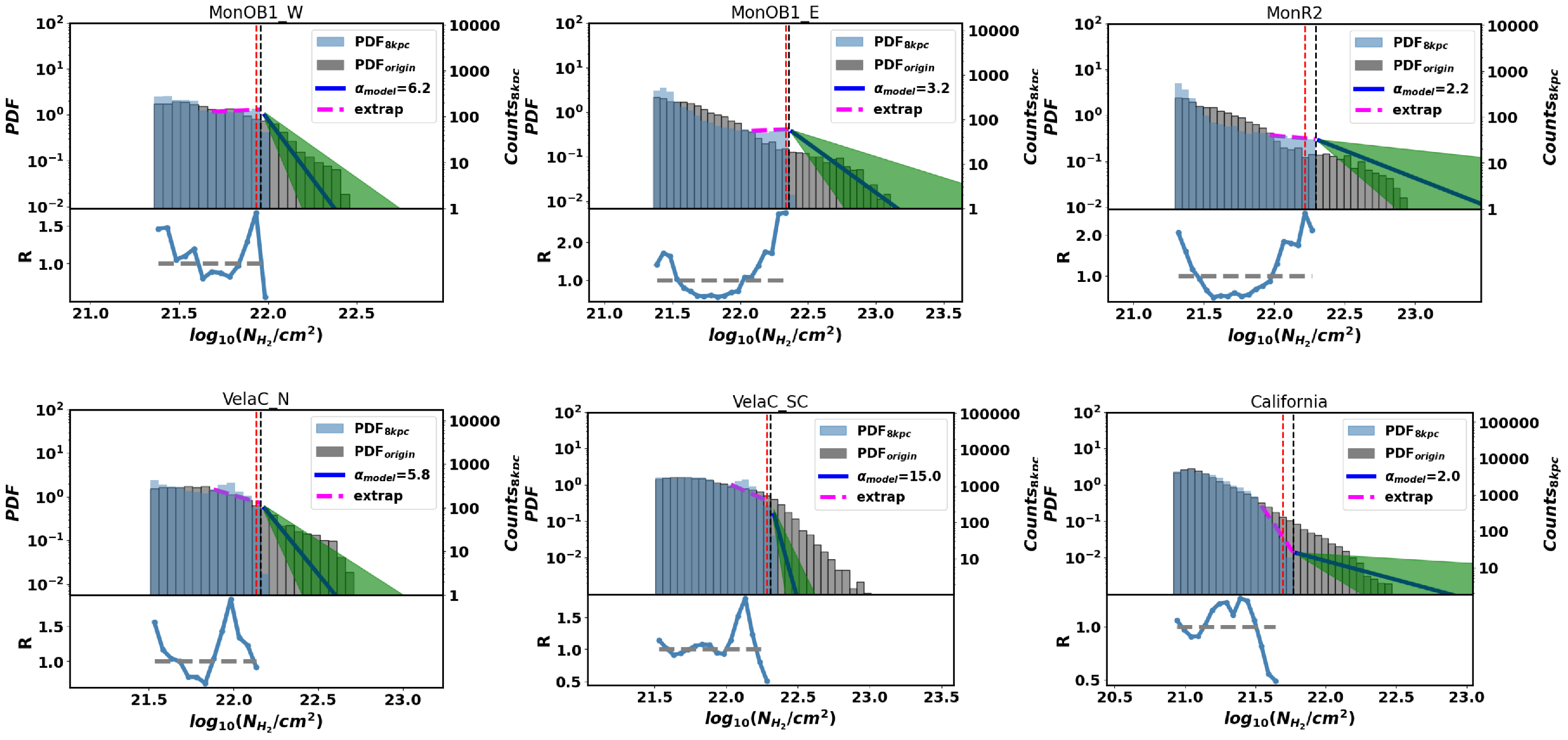}
\caption{Same as Figure~\ref{fig:2kpc}, but for simulated observations at 8 kpc.}
\label{fig:8kpc}
\end{figure*}

Figures~\ref{fig:2kpc} and~\ref{fig:8kpc} illustrate the best inferred power-law fits for the high-density tails at simulated distances of 2 kpc and 8 kpc, respectively. The column density ranges used for extrapolation are indicated by magenta lines. This range is determined uniformly across the sample as follows: (1) the cumulative probability of the low-resolution N-PDF is calculated from the normalized histograms of each region; (2) the second derivative of the cumulative probability is computed; and (3) the column densities corresponding to the first and second points of maximum curvature (where the second derivative changes sign) in the cumulative N-PDF are selected as the endpoints for the extrapolation. Here, ``first'' and ``second'' refer to the order starting from the largest bin with non-zero probability. If the second derivative does not change sign up to 0.3 dex below the bin of the first maximum curvature, then the range between the bin of the first maximum curvature and the bin at $  -0.3  $ dex is used for the extrapolation.

\begin{figure}[!htbp]
\centering
\includegraphics[scale=0.58]{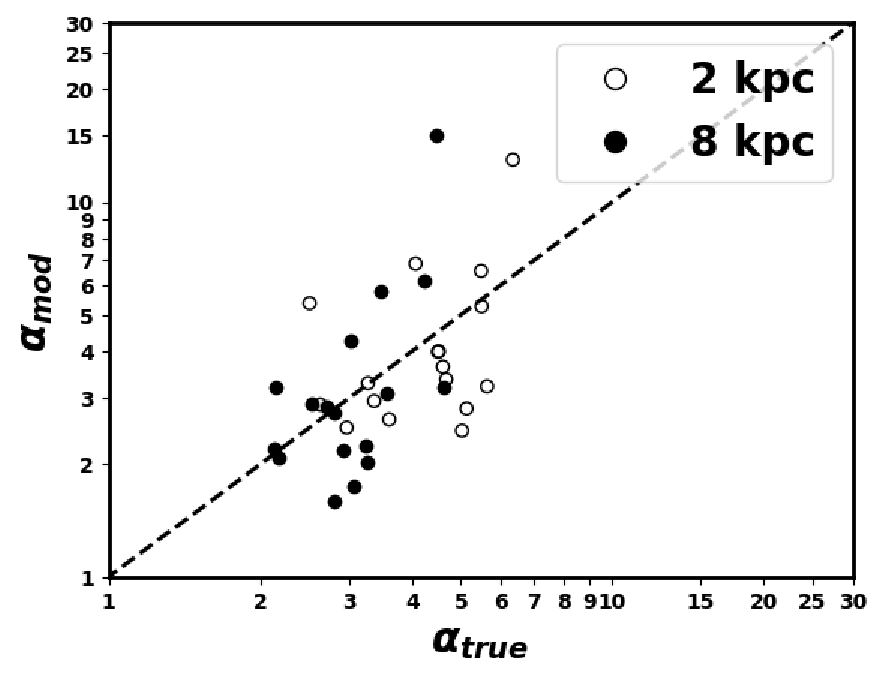}
\caption{Comparison between the recovered SFEq power-law index ($  \alpha_{\rm mod}  $) and the true SFEq index ($  \alpha_{\rm true}  $).} 
\label{fig:alpha}
\end{figure}

We find a relatively constant scatter of $  \approx 50\%  $ when comparing the recovered SFEq power-law index to the intrinsic SFEq power-law index at high resolution (Figure~\ref{fig:alpha}). This uncertainty is plotted as a shaded area in Figures~\ref{fig:2kpc} and~\ref{fig:8kpc}. This simple prediction does not perfectly recover the intrinsic power-law in all regions. First, we identify a subsample of regions that exhibit a gradual decline above $  N_{\rm trun}  $, leading to underestimated $  P_{\rm high}(\log_{10}N_{\rm trun})  $ via extrapolation and consequently low $  \alpha_{\rm mod} \lesssim 2.2  $; examples include California, Orion A, Perseus West, and IC 5146 at 8 kpc, which also show overestimated truncation column densities by the proportional nesting model. Second, the low-resolution N-PDFs commonly display a U-shaped distortion relative to their high-resolution counterparts, as illustrated in the lower panels of Figures~\ref{fig:2kpc} and~\ref{fig:8kpc}. Such distortion is naturally expected from smearing effects, which cause accumulations at both the high and low ends of the PDFs. The distortion is found to be stronger at larger distances (or lower resolutions) and is most pronounced in Mon R2 and Mon OB1 East at large distances, leading to overestimated $P_{\rm high}(N_{\rm trun})$ and consequently higher $  \alpha_{\rm mod} $. Correspondingly, as discussed previously, these two regions also exhibit the flattest power-law tails and have truncation column densities that are underestimated by the proportional nesting model.

\section{Discussion} \label{sec:discuss}

\subsection{SFEq power-law index}

Throughout this study, we derive effective power-law indices by equating the integral of a power-law distribution to the observed survival function $P(>N)$, assuming that we are only studying the power-law tails of the N-PDFs. A key advantage of this approach over traditional Maximum Likelihood Estimation (MLE) for fitting power-law indices in N-PDFs of molecular clouds is its simplicity and applicability to complex N-PDFs featuring multiple sections of power-law tails, as it aggregates the cumulative behavior without requiring segmented fits or assumptions about break points, making it robust for structures where tails may deviate gradually or exhibit bumps. In reality, N-PDFs of molecular clouds rarely follow a single power-law distribution, while MLE provides statistically robust estimates assuming a Pareto-like form, for complex N-PDFs, it requires extensions like segmented MLE or minimum $N_{\rm thresh}$ optimization to handle breaks/multiple sections, which can introduce subjectivity in defining ranges. The only caveat is that the SFEq index inherently weights the high-end tail more strongly compared to the balanced MLE estimator. The survival function $  P(>N)  $ accumulates contributions cumulatively from the extremes, so rare,high-$  N  $ events (where $  P(>N)  $ is small but the slope deviation is pronounced) disproportionately shape the effective $  \alpha  $—essentially, the offsets in the high density tail propagate back through the integral. For complex N-PDFs, this makes the index robust to fluctuations in the body near $  N_{\rm thresh}  $ but also sensitive to truncation or outliers at the upper end.

\subsection{The Link between N-PDFs and the morphologies of molecular clouds}

The N-PDFs of molecular clouds serve as a reasonable diagnostic tool for probing the underlying morphologies and physical processes governing their structure, particularly in the context of turbulence, gravity, and stellar feedback. Our findings reveal that regions exhibiting flatter power-law tails in their N-PDFs demonstrate greater resistance to beam smearing effects, suggesting a morphology characterized by more extended, coherent dense structures rather than the compact, hierarchically embedded clumps implied by a pure proportional nesting model. This resilience manifests as a delayed low-resolution-induced truncation in N-PDFs, where smearing dilutes high-density contrasts less aggressively due to the broader spatial distribution of dense gas, potentially driven by large-scale compression or feedback that preserves filamentary networks over isolated peaks.

In this study, whenever the intrinsic SFEq power-law index is smaller than 2.2, the difference $\log_{10}(N_{\rm trun,mod}/N_{\rm trun,sim})$ is always greater than $-0.15$ dex (i.e., the proportional nesting model underestimates the truncation density by a relatively small amount). We therefore consider an intrinsic power-law index $\alpha_{\rm SFEq} \lesssim 2.2$ to represent a ``flatter'' tail in the context of this study. This threshold is somewhat lower than many typical values reported in observations and simulations of evolved molecular clouds (often $\alpha \approx 2.0$--3.0 or steeper; e.g., Girichidis et al. 2014; Schneider et al. 2015, 2022). Stronger constraints on this threshold could be obtained by expanding the sample with additional high-resolution observations or numerical simulations.

These observations align closely with \citet{schneider22}, who found that the characteristic sizes of molecular clouds correlate with the transition column density of the second power-law tail, proposing that stellar feedback and magnetic fields alter N-PDFs by reshaping cloud morphologies---e.g., compressing gas into denser shells or pillars, which flattens tails and shifts transition points to higher densities. Specifically, they argue that feedback slows down collapse and reduces mass flow toward higher densities, leading to flatter power-law tails. This is supported by \citet{tremblin14}, who find flatter power-law tails near H\,{\sc ii} interaction zones due to compression. Furthermore, they argue that clouds/regions with a second steeper power-law tail show a perpendicular relative orientation between the density structures and the magnetic field, based on comparisons of the N-PDFs with those in \citet{soler19}.

On the other hand, regions with pronounced multiplicity display overestimated truncation densities $N_{\rm trun}$ and sharper truncations in low-resolution N-PDFs, indicating that disconnected substructures (e.g., fragmented filaments) accelerate tail suppression beyond model predictions.

Linking observed N-PDF shapes---especially high density tails---to these morphological interpretations in practice is complicated by the fact that real observations always suffer from beam-smearing effects. We suggest using the first point of maximum curvature in the cumulative distribution (as defined in Section~4) as the reliable high-density boundary when fitting the apparent power-law tail. The low-density boundary of the power-law regime remains more ambiguous, as it depends on the chosen decomposition method and possible foreground/background fluctuations. Our simulations do not yield a single universal minimal span in column density---whether expressed in dex above $N_{\rm trun}$ or relative to beam size---for which smearing effects on the measured slope can be safely neglected. A more detailed quantification of the minimum reliable tail span and the precise range of slopes for which smearing effects can be neglected is left for future work.

\subsection{The effects of resolution on N-PDF decomposition}

Studies of N-PDFs often focus on separating the log-normal portion (dominating the low-density, turbulence-driven regime) of the N-PDFs and the power-law tails (reflecting gravity-influenced high-density collapse). The decomposition of low-resolution N-PDFs can be significantly compromised by the dual effects of truncation at high column densities and U-shape distortions below this threshold, leading to biased parameter estimates and potential misinterpretations of cloud physics. The truncation, arising from beam smearing that dilutes compact high-N peaks into unresolved averages, shortens the observable power-law tail, often resulting in an artificially steeper fitted slope or an underestimated transition density where the power-law begins, as the fit struggles to capture the abrupt cutoff without overcompensating by shifting the break point lower or inflating the log-normal width to absorb the suppressed high end. Meanwhile, the U-shaped distortion—characterized by boosted probabilities at both the low-N end (due to artificial low-density excesses from averaging voids or diffuse gas) and near the truncation (from normalized redistribution of smeared high-N contributions)—further warps the log-normal core, broadening its variance or introducing spurious peaks that mimic multi-component turbulence or feedback effects, thereby inflating the fitted log-normal width ($  \sigma  $) or underestimating the power-law contribution's mass fraction. Collectively, these biases can obscure genuine morphological insights, such as flatter tails from feedback-compressed extended structures, emphasizing the need for resolution-corrected models to reliably disentangle intrinsic cloud properties from observational artifacts in N-PDF decompositions. 

Our reverse-modeling method recovers the intrinsic high-density power-law tail with a typical uncertainty of approximately 50\%. This uncertainty primarily arises from two sources: (1) limitations inherent to the proportional nesting model, including the effects of multiplicity and tail flattening, and (2) our use of the apparent point of maximum curvature in the observed cumulative distribution as a proxy for the upper end of the intrinsic tail. Nevertheless, the method is already directly applicable to distant high-mass star-forming regions such as those studied by Schneider et al.\ (2022). For example, Cygnus North, M17, and NGC 6334 lie at distances of 1.3--2.2\,kpc, which matches the regime explored in our 2\,kpc simulations. These regions exhibit notably flat power-law tails. In such cases, recovering an SFEq index $\alpha_{\rm SFEq} \lesssim 2.5$ together with a relatively flat apparent slope just below the truncation density would provide strong evidence for an intrinsically flat power-law tail that extends beyond the resolution limit, rather than being an artifact of severe smearing.

\section{Summary} \label{sec:summary}

In this work, we have investigated the effects of observational resolution on the N-PDFs of molecular clouds, focusing on the truncation of their high-density power-law tails. Using a sample of 17 local massive molecular clouds with high-resolution Herschel-derived column density maps, we simulated low-resolution observations at distances from 2 to 10 kpc to mimic the challenges faced in studying distant clouds.

\begin{itemize}

\item We introduced a simple proportional nesting model, which assumes that high-density regions are hierarchically embedded within lower-density gas in a self-similar manner. This model predicts the truncation column density $  N_{\rm trun}  $ as the point where the cumulative probability $  P(>N_{\rm trun})  $ equals the ratio of the beam area to the threshold-enclosed area. Comparisons with simulated data show remarkable agreement, with deviations within 0.3 dex, highlighting the model's robustness. Outliers are explained by multiplicity of semi-independent or disconnected dense substructures in large complexes (leading to overestimations) and flatter power-law tails (leading to underestimations), with the characteristic cloud scale $  L_{\rm char}  $, derived from $  \Delta  $-variance spectra, serving as a key indicator of when global smearing effects dominate.

\item Furthermore, we developed a method to reconstruct the missing high-density N-PDF tail from low-resolution observations by extrapolating a power-law fit and adjusting for the predicted missing probability, incorporating a filling-factor correction. This approach recovers intrinsic power-law indices with an uncertainty of approximately $50\%$, though it is susceptible to biases from gradual declines or U-shaped distortions in certain regions.

\item Our findings link N-PDF shapes to underlying cloud morphologies: flatter tails indicate extended, coherent structures resistant to smearing, potentially shaped by stellar feedback or compression, while multiplicity accelerates truncation. We also discussed the advantages of the survival-function-equivalent (SFEq) power-law index for analyzing complex N-PDFs and cautioned that resolution effects can bias decompositions into log-normal and power-law components, inflating variances or steepening slopes.

\end{itemize}

Overall, this study provides a framework for correcting resolution biases in N-PDF analyses, enabling more accurate constraints on turbulent star formation models in distant systems. Future work could extend the model to incorporate 3D simulations, magnetic fields, or stellar feedback to refine predictions for diverse cloud environments.

\begin{acknowledgements}

We thank the referee for the constructive comments and suggestions which helped us to improve the manuscript. We acknowledge support from the National Natural Science Foundation of China (grant No. 12141302). This work is also supported in part by the Chinese Academy of Sciences (CAS) through a grant to the CAS South America Center for Astronomy (CASSACA) in Santiago, Chile.

\end{acknowledgements}

\bibliographystyle{aa}
\bibliography{references}

\begin{appendix}

\onecolumn  

\section{Complete figures for the entire sample}

\begin{figure*}[!ht]
\centering
\includegraphics[scale=0.3]{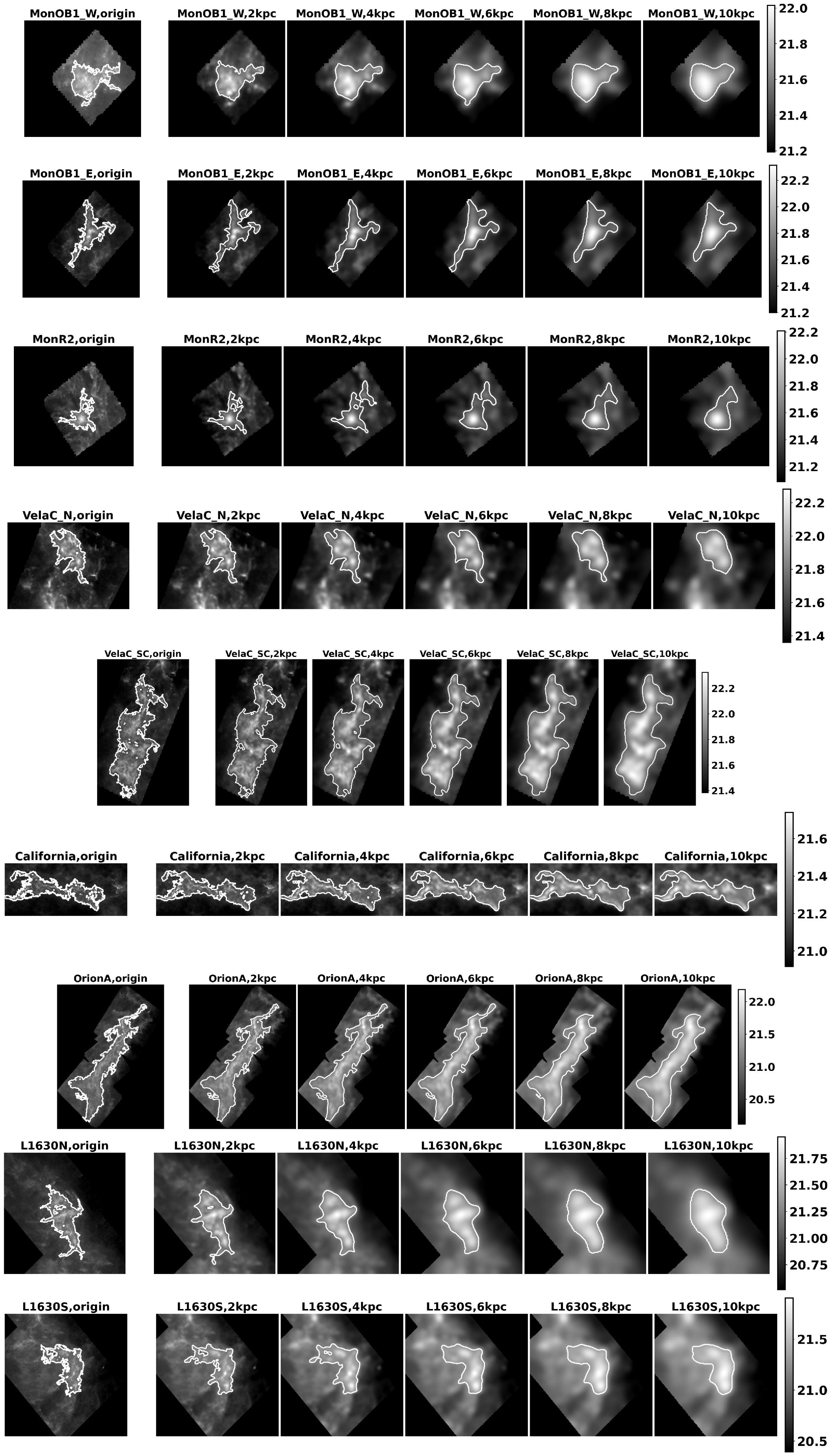}
\caption{Complete set of Figure~\ref{fig:simu} for the entire sample. (a)}
\end{figure*}

\begin{figure*}[!ht]
\centering
\includegraphics[scale=0.3]{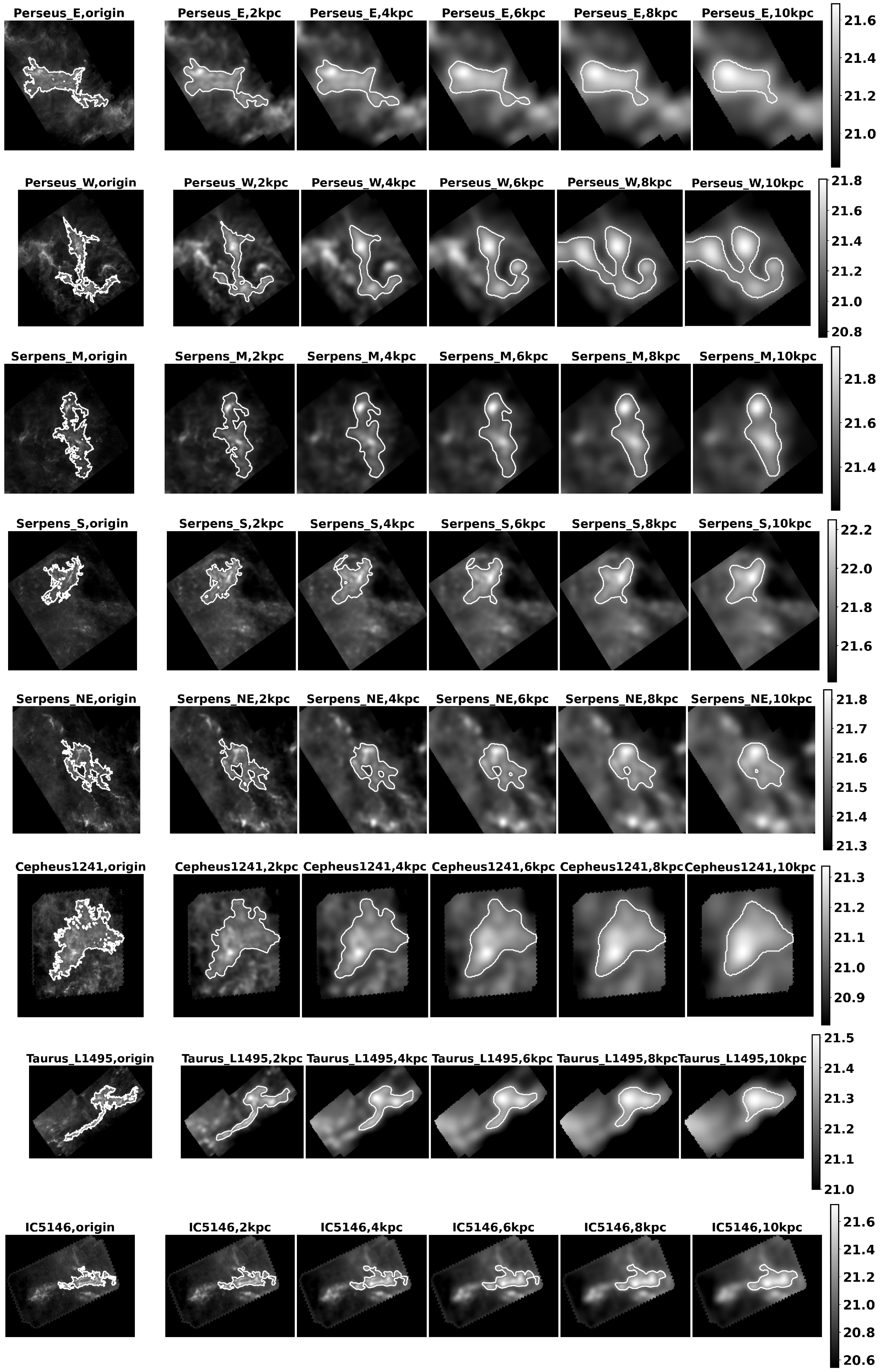}
\caption{Complete set of Figure~\ref{fig:simu} for the entire sample. (b) Continued.}
\end{figure*}

\begin{figure*}[!ht]
\centering
\includegraphics[scale=0.7]{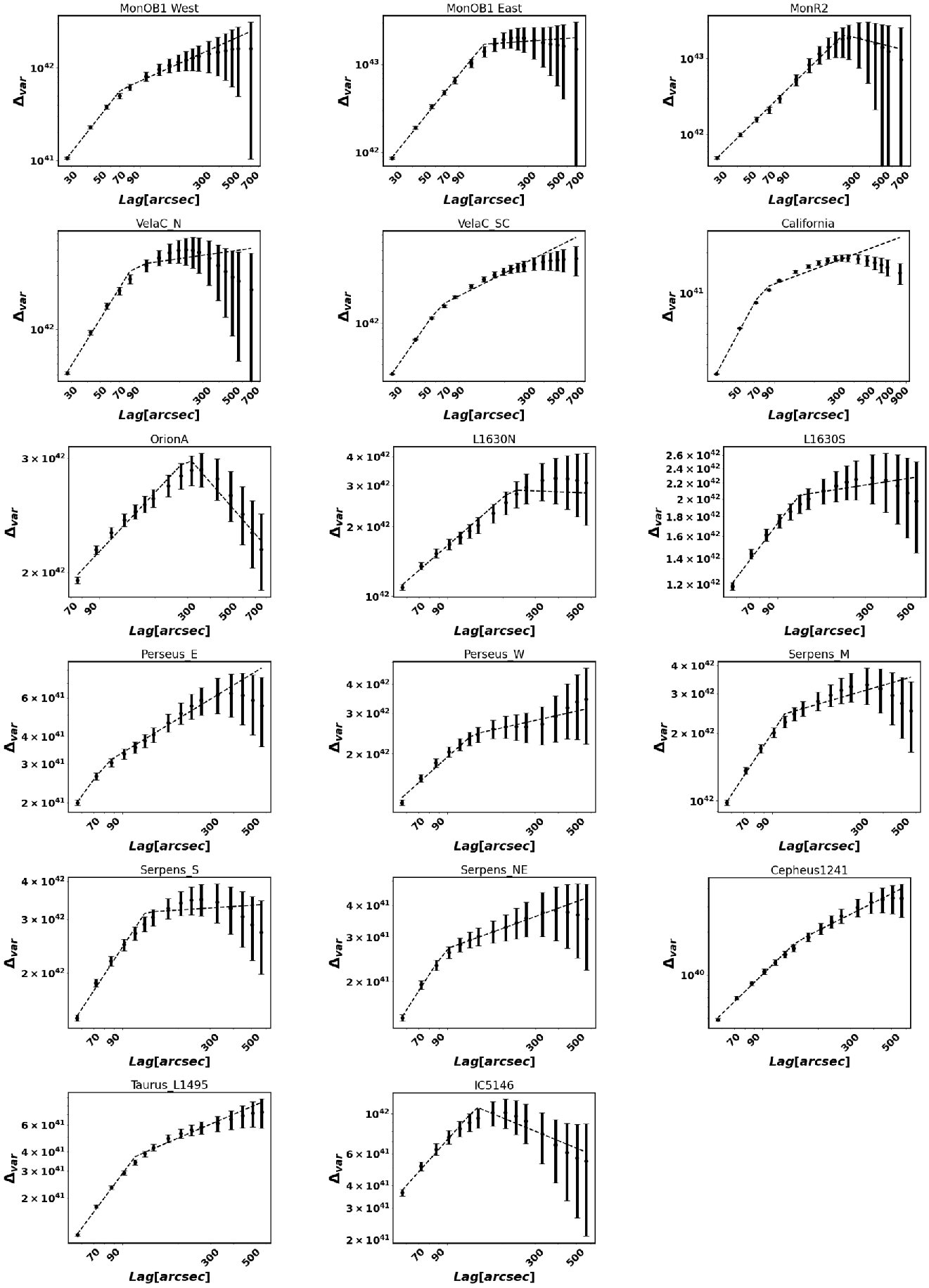}
\caption{Complete set of Figure~\ref{fig:deltav} for the entire sample.}
\end{figure*}

\begin{figure*}[!ht]
\centering
\includegraphics[scale=0.58]{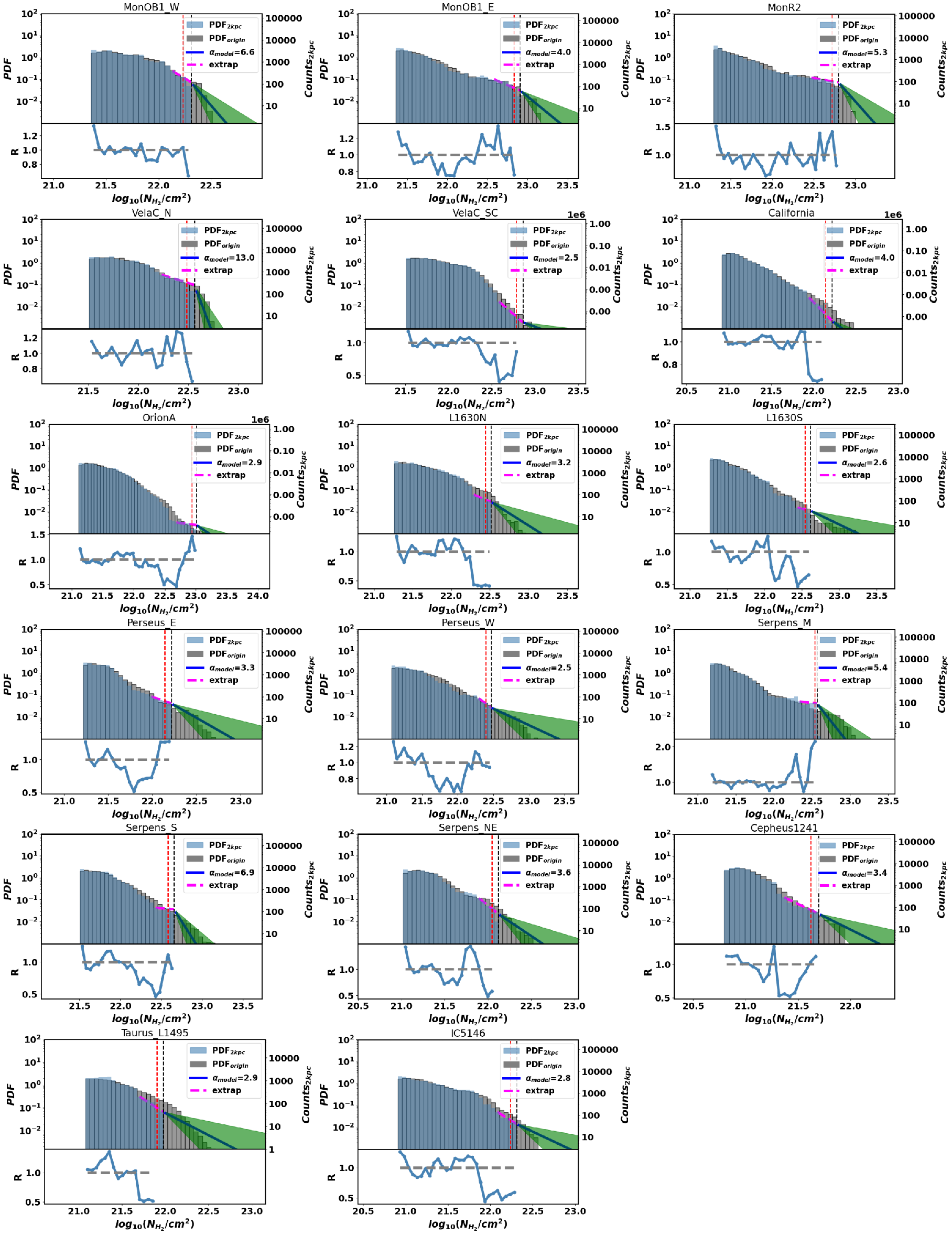}
\caption{Complete set of Figure~\ref{fig:2kpc} for the entire sample.}
\end{figure*}

\begin{figure*}[!ht]
\centering
\includegraphics[scale=0.58]{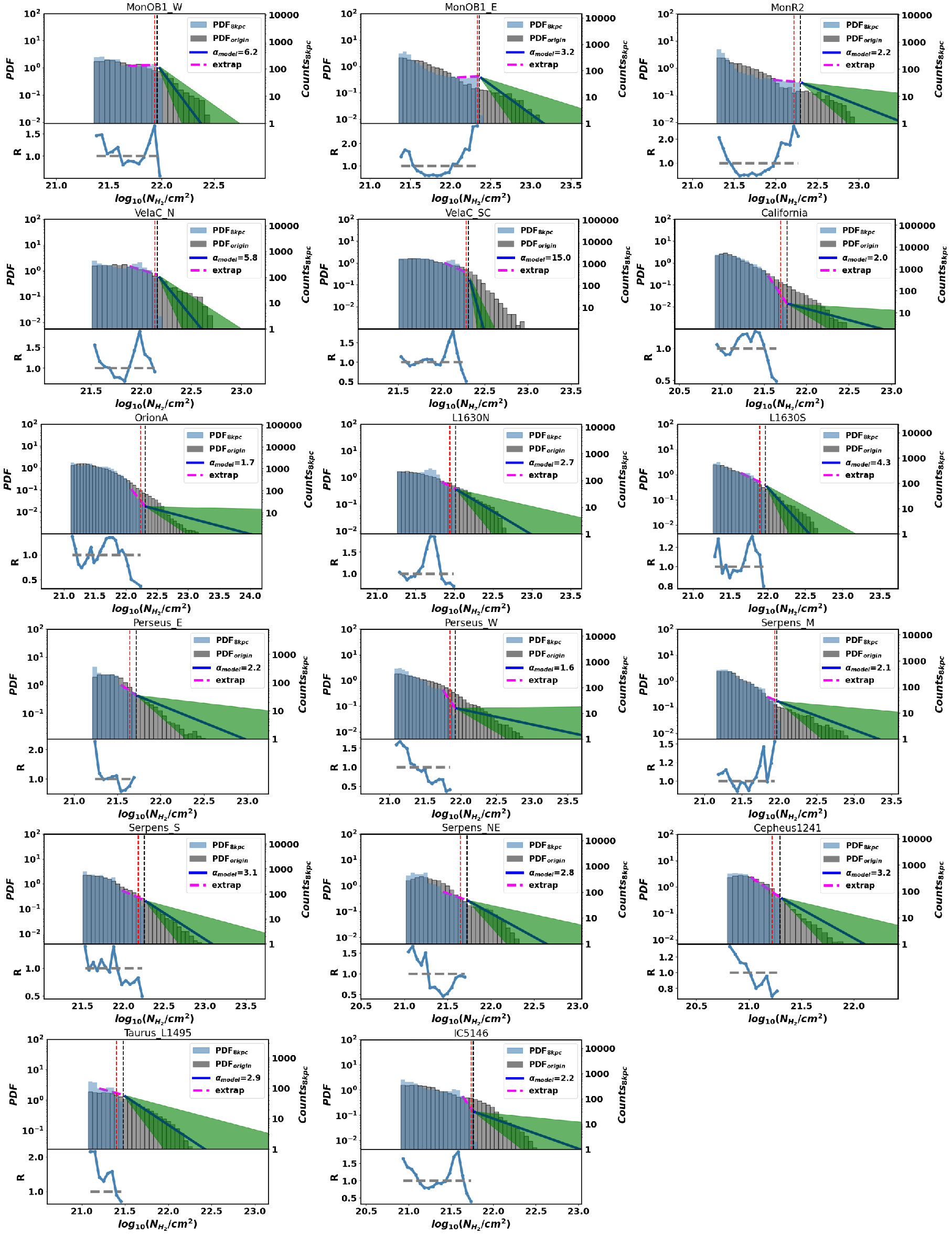}
\caption{Complete set of Figure~\ref{fig:8kpc} for the entire sample.}
\label{fig:8kpc_app}  % optional distinct label if needed
\end{figure*}

\twocolumn
\end{appendix}

\end{document}